\documentclass{aipproc}

\layoutstyle{8x11single}

\begin{document}

\title{Updated Search for Electron Antineutrino Appearance at MiniBooNE}

\classification{14.60.Pq,14.60.St}
\keywords      {Antineutrino oscillations, LSND, MiniBooNE}

\author{E.\ D.\ Zimmerman}{
  address={University of Colorado, Boulder, Colorado 80309\\{\rm For the MiniBooNE Collaboration} }
}

\begin{abstract}
The MiniBooNE experiment at Fermilab has updated its search for $\bar\nu_\mu \rightarrow \bar\nu_e$ oscillations with data collected through May 2011.  This represents a statistics increase of 52\% over the result published in 2010.  The data favor LSND-like oscillations over a background-only hypothesis at the 91.1\% confidence level.  While the new result remains equally consistent with LSND, the compatibility with the background-only hypothesis is improved.  An excess of $38.6 \pm 18.5$ $\nu_e$-like events below 475~MeV is observed, consistent with the observation of such an excess in neutrino mode.
\end{abstract}

\maketitle


\section{MiniBooNE}

The MiniBooNE experiment (E898/944) at Fermi National Accelerator
Laboratory is a short-baseline neutrino oscillation experiment whose
main purpose is to test the LSND oscillation results \cite{lsnd}.
MiniBooNE uses an 8~GeV proton beam from the Fermilab Booster to
produce pions, which then decay in flight to produce a nearly pure
$\nu_\mu$ flux at a mineral oil Cherenkov detector 500~m away. The
detector uses Cherenkov ring shape information to distinguish
charged-current $\nu_\mu (\bar\nu_\mu)$ from $\nu_e (\bar\nu_e)$
interactions, searching for an excess of $\nu_e$ which would indicate
oscillations. Data collection began in late 2002 with the beam
configured to produce neutrinos; since 2007 most operations have been
with the focusing polarity reversed to produce antineutrinos.  Antineutrino
studies are important for a complete test of LSND, which was primarily
a $\bar\nu_\mu \rightarrow \bar\nu_e$ search.

\section{Oscillation Results from MiniBooNE}

MiniBooNE has three previous appearance-mode oscillation results.  The
general technique is similar in all of them: events with a single
electron-like Cherenkov ring are selected as charged-current
quasielastic (CCQE) candidates.  Cuts are designed to remove
neutral-current $\pi^0$ events as well as fragments from neutrino
interactions that occurred outside the detector (``dirt'').  The
reconstructed neutrino energy $E_\nu^{\rm QE}$ is computed assuming
the process was $\nu N \rightarrow e N^\prime$ with the nucleon unobserved and the neutrino
originating from the beam direction.  Significant backgrounds are
intrinsic $\nu_e$ in the beam, neutral-current $\pi^0$ and $\Delta
\rightarrow N\gamma$, and dirt. Neutrinos and antineutrinos cannot be
distinguished on an event-by-event basis.

In 2007, the main search for LSND-like oscillations in neutrino mode was
published \cite{boone-nu} based on $5.7\times 10^{20}$ protons on
target (POT).  This result excluded a $CP$-conserving two-neutrino
oscillation explanation for LSND at the 98\% confidence level.  At the
same time, an unexplained excess of $\nu_e$-like events below 475~MeV
was seen; further studies of this low-energy excess were published in
2009 \cite{lowe-prl}.  

In 2010, a search for $\bar\nu_\mu \rightarrow \bar\nu_e$ from
$5.66\times 10^{20}$ POT in antineutrino running mode was published
\cite{boone-nubar10}.  That paper indicated an excess of $20.9 \pm
14.0$ $\nu_e/\bar\nu_e$ candidates in the 475-1250~MeV range where MiniBooNE
is most sensitive to LSND-like oscillations and contributions from
the neutrino-mode low-energy excess are minimized.  However, a
likelihood-ratio fit to the energy distributions of both the
$\bar\nu_e$ and $\bar\nu_\mu$ candidates preferred the oscillation
hypothesis over the background-only hypothesis with 99.4\%
probability.  In the low-energy region of 200-475~MeV, the data excess
over the background prediction was $18.5 \pm 14.3$ events.  Scaling the
neutrino-mode low-energy excess by the expected neutrino contamination
in the antineutrino flux gives a predicted low-energy excess of 12 events.

\section{UPDATED ANTINEUTRINO RESULTS}

In this section we present preliminary updates to the antineutrino
oscillation search reported in Ref.~\cite{boone-nubar10} using
$8.58\times 10^{20}$ POT, using a nearly unchanged analysis.  The most
significant change is an improved constraint on kaon-decay $\nu_e$ in
the beam from the SciBooNE experiment \cite{sciboo-k}, reducing the
prediction for that background source by 3\% and its error by a factor
of three.  Detector and beam monitoring have indicated no significant
changes over the entire run period, and Kolmogorov-Smirnov tests of
neutrino control samples are consistent with a constant event rate.

\begin{figure}
  \includegraphics[height=.39\textheight]{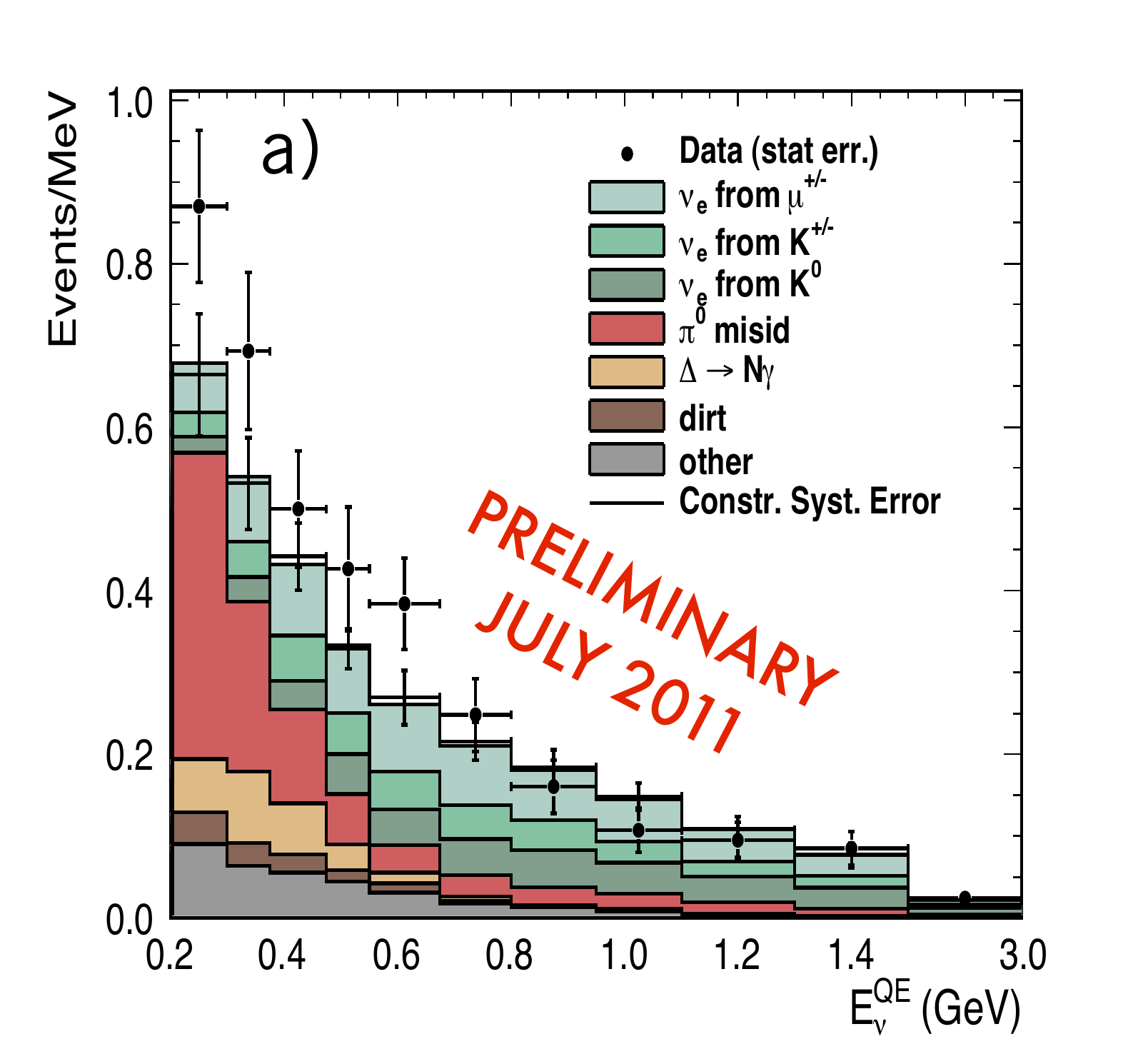}
  \includegraphics[height=.37\textheight]{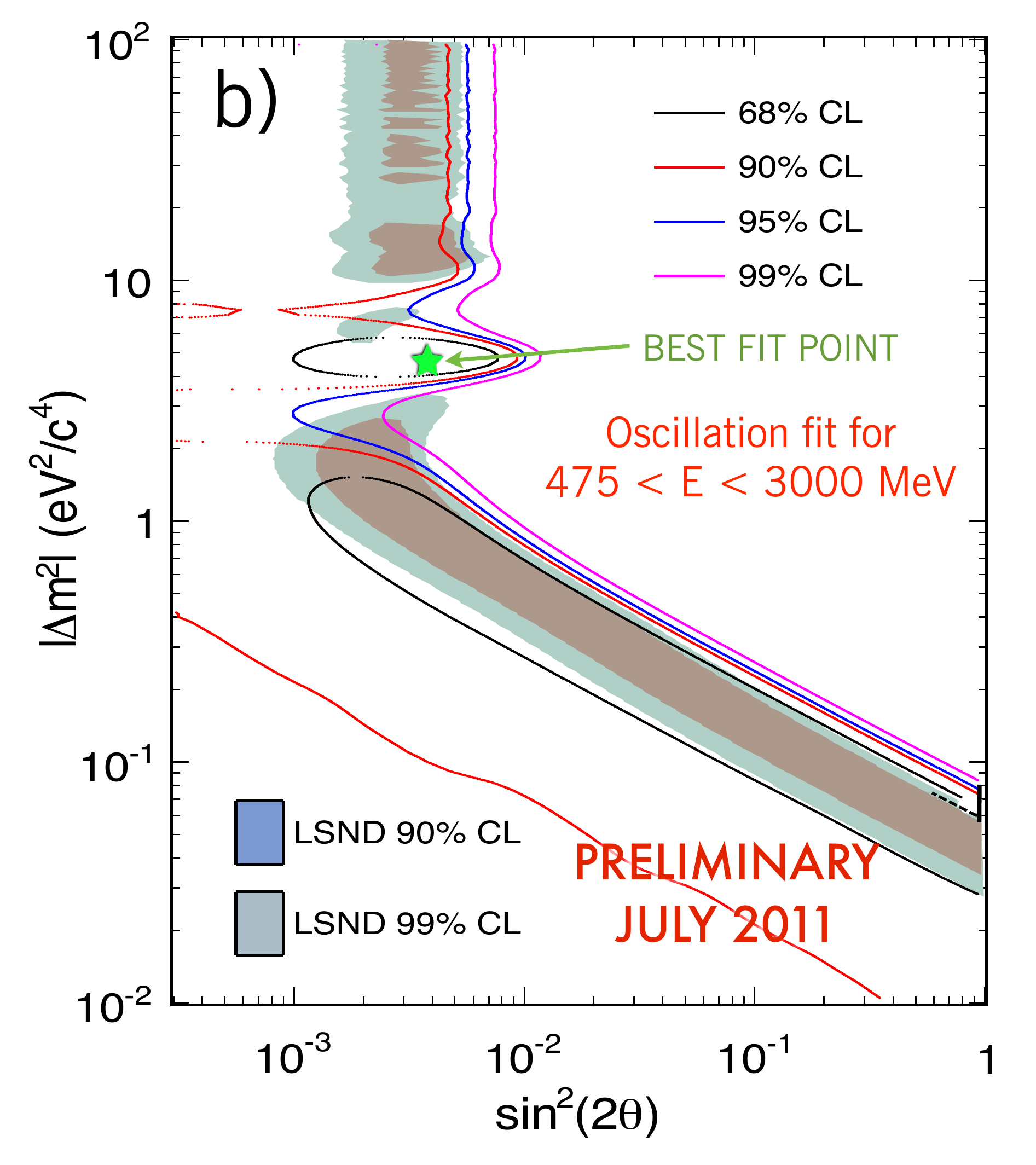}
  \caption{{\bf a)}~Reconstructed $E_\nu^{\rm QE}$ distribution for $\bar\nu_e$
    candidates with data collected in antineutrino mode through May
    2011.  
    {\bf b)}~Oscillation fit contours for updated $\bar\nu_e$ appearance
    analysis.  Energy range of fit is $475<E_\nu^{\rm QE}<3000$~MeV.
    Results are preliminary. \label{enuqe-distr}}
\end{figure}

The reconstructed energy distribution
of $\bar\nu_e$ candidates is shown in Fig.~\ref{enuqe-distr}a.  
In the augmented data set, 168 events are observed in the 475-1250~MeV
region, corresponding to an excess of $16.3\pm 19.4$ over predicted
background.  The excess in this region is thus reduced when the new data
are added.  

In the low-energy region, the excess has grown to $38.6\pm 18.5$
events.  We can compare this excess to predictions from scaling the
observed neutrino-mode low-energy excess by various factors (see
Table~\ref{scaling}).  This can exclude some possible phenomena as primary sources
of the low-energy excess, though it should be noted that none of these
processes can be scaled by a large enough factor to explain the low-energy
excess without disagreeing
strongly with control samples.
\begin{table}
\begin{tabular}{lcclc}
\hline
  \tablehead{1}{c}{b}{Scaling basis}
  & \tablehead{1}{c}{b}{Predicted excess}
  & \tablehead{1}{c}{b}{\hspace{0.0in}}
  & \tablehead{1}{c}{b}{Scaling basis}
  & \tablehead{1}{c}{b}{Predicted excess}
\\
\hline
Total background & 50 & \hspace{0.4in} & Neutrino contamination & 17 \\
$\Delta \rightarrow N\gamma$ decays & 39 & & Dirt & 46 \\
Protons on target\tablenote{Excess would scale this way if due to neutral particles in secondary beam.}& 165 & & $K^+$ in secondary beam & 67 \\
Neutral-current $\pi^0$ & 48 & & Inclusive charged-current & 59\\
\hline
\end{tabular}
\caption{Preliminary predicted low-energy excess from scaling neutrino-mode excess by various factors.  The value observed in data is $38.6\pm 18.5$.  }
\label{scaling}
\end{table}

The primary test of LSND's result is the simultaneous fit to the
$\bar\nu_\mu$ and $\bar\nu_e$ candidates in the $475<E_\nu^{\rm QE}<3000$~MeV range.  The updated
confidence-level contours are shown in Fig.~\ref{enuqe-distr}b.  The
fit prefers the oscillation hypothesis to the background-only
hypothesis at 91.1\% confidence level.  The best-fit point moved from
the high-$\sin^2(2\theta)$, low-$\Delta m^2$ region to the
high-$\Delta m^2$ ``island'' solution, however the $\chi^2$ minimum is
quite broad so this does not represent a significant change.  In the
signal bins, the $\chi^2$ probability is 14.9\% for background-only
and 35.5\% for the oscillation fit. (These numbers were 0.5\% and 10\%
before the statistics update.)  The 68\% confidence level contour
still covers most of the LSND allowed region.  This result is
therefore still consistent with LSND, but the evidence for LSND-like
oscillations is no longer as strong.

As in the published result, fits have been performed under other
sets of assumptions.  First, a fit to the entire energy range ($200<
E_\nu^{\rm QE}<3000$~MeV) yields similar contours, with oscillations
preferred with 97.6\% probability.  The fit $\chi^2$ probabilities for
background-only and the best oscillation fit are 10.1\% and 50.7\%
respectively.  However, it should be noted that there is a large known
neutrino contamination in antineutrino running (22\% of the
$\bar\nu_\mu$ candidates and 44\% of the total background to the
$\bar\nu_e$ oscillation signal).  A low-energy excess from neutrinos
may be expected to contribute to the 200-475~MeV
bins in this fit, but has not been subtracted because its origin and scaling
are unknown.  Therefore, this fit cannot be interpreted as a pure antineutrino
physics result.

Another alternative fit is done with additional background subtraction, assuming
that the low-energy excess contribution from neutrinos simply scales
with the neutrino-induced event rate in each bin.  This model's validity may
be poor, however, if the excess is due to feed-down from misreconstructed
higher-energy neutrinos with unobserved particles in the final state (since
the neutrino spectrum differs between neutrino and antineutrino running).
The fit under this background subtraction model prefers oscillations with
94.2\% probability.  The fit $\chi^2$ probabilities for
background-only and the best oscillation fit are 28.3\% and 76.5\%
respectively.  

The contours and best-fit points for the fits to the full energy spectrum
are available in the slides from this presentation.


\section{Conclusion and next steps}

MiniBooNE presents a preliminary update to its 2010 electron
antineutrino appearance search with 52\% more integrated flux.  Adding
the new data reduces the significance of the apparent LSND-like
oscillation signal, and reveals a low-energy excess of $\bar\nu_e$
candidates similar to that observed in neutrino mode.  In the
higher-energy region, the new data are consistent with both LSND
and a background-only hypothesis.

The experiment will continue to take data at least until the 2012
shutdown.  The collaboration's goal is a total of $1.5\times 10^{21}$
POT -- nearly doubling the current data set.  This
statistically-limited result should benefit greatly from the
additional data.  If either our best-fit value or LSND's central value
is correct, the expected exclusion of the background-only hypothesis
(relative to the oscillation fit) with a data set that large is at the 98-99\% confidence level.



\begin{theacknowledgments}
 The author and the MiniBooNE collaboration acknowledge the contributions of
 Fermi National Accelerator laboratory and support from 
 the Department of Energy and the National Science Foundation.
\end{theacknowledgments}



\bibliographystyle{aipprocl} 




\end{document}